\documentclass[twocolumn,english,aps,twocolum,superscriptaddress,amsmath,amssymb,floatfix]{revtex4-1}
\usepackage[colorlinks=true,citecolor=blue,linkcolor=magenta]{hyperref}

\usepackage[markup=blue, authormarkupposition=left]{changes} 

\usepackage{soul}
\usepackage[utf8]{inputenc}
\usepackage[english]{babel}
\usepackage{amsmath,amsfonts,amssymb}
\usepackage[T1]{fontenc}
\usepackage{url}

\usepackage{amsmath}
\usepackage{amsfonts}
\usepackage{amssymb}

\usepackage{epstopdf}
\usepackage{graphicx}

\begin{document}

\title{Photonic chip-based low noise microwave oscillator}

\author{Igor Kudelin}
\affiliation{National Institute of Standards and Technology, 325 Broadway, Boulder, CO 80305, USA}
\affiliation{Department of Physics, University of Colorado Boulder, 440 UCB Boulder, CO 80309, USA}

\author{William Groman}
\affiliation{National Institute of Standards and Technology, 325 Broadway, Boulder, CO 80305, USA}
\affiliation{Department of Physics, University of Colorado Boulder, 440 UCB Boulder, CO 80309, USA}

\author{Qing-Xin Ji}
\affiliation{T. J. Watson Laboratory of Applied Physics, California Institute of Technology, Pasadena, CA 91125, USA}

\author{Joel Guo}
\affiliation{Department of Electrical and Computer Engineering, University of California, Santa Barbara, Santa Barbara, California 93106, USA}

\author{Megan L. Kelleher}
\affiliation{National Institute of Standards and Technology, 325 Broadway, Boulder, CO 80305, USA}
\affiliation{Department of Physics, University of Colorado Boulder, 440 UCB Boulder, CO 80309, USA}

\author{Dahyeon Lee}
\affiliation{National Institute of Standards and Technology, 325 Broadway, Boulder, CO 80305, USA}
\affiliation{Department of Physics, University of Colorado Boulder, 440 UCB Boulder, CO 80309, USA}

\author{Takuma Nakamura}
\affiliation{National Institute of Standards and Technology, 325 Broadway, Boulder, CO 80305, USA}
\affiliation{Department of Physics, University of Colorado Boulder, 440 UCB Boulder, CO 80309, USA}

\author{Charles A. McLemore}
\affiliation{National Institute of Standards and Technology, 325 Broadway, Boulder, CO 80305, USA}
\affiliation{Department of Physics, University of Colorado Boulder, 440 UCB Boulder, CO 80309, USA}

\author{Pedram Shirmohammadi}
\affiliation{Department of Electrical and Computer Engineering, University of Virginia, Charlottesville, VA 22904, USA}

\author{Samin Hanifi}
\affiliation{Department of Electrical and Computer Engineering, University of Virginia, Charlottesville, VA 22904, USA}

\author{Haotian Cheng}
\affiliation{Department of Applied Physics, Yale University, New Haven, Connecticut 06520, USA}

\author{Naijun Jin}
\affiliation{Department of Applied Physics, Yale University, New Haven, Connecticut 06520, USA}

\author{Sam Halliday}
\affiliation{Department of Applied Physics, Yale University, New Haven, Connecticut 06520, USA}

\author{Zhaowei Dai}
\affiliation{Department of Applied Physics, Yale University, New Haven, Connecticut 06520, USA}

\author{Lue Wu}
\affiliation{T. J. Watson Laboratory of Applied Physics, California Institute of Technology, Pasadena, CA 91125, USA}

\author{Warren Jin}
\affiliation{Department of Electrical and Computer Engineering, University of California, Santa Barbara, Santa Barbara, California 93106, USA}

\author{Yifan Liu}
\affiliation{National Institute of Standards and Technology, 325 Broadway, Boulder, CO 80305, USA}
\affiliation{Department of Physics, University of Colorado Boulder, 440 UCB Boulder, CO 80309, USA}

\author{Wei Zhang}
\affiliation{Jet Propulsion Laboratory, California Institute of Technology, Pasadena, California, USA}

\author{Chao Xiang}
\affiliation{Department of Electrical and Computer Engineering, University of California, Santa Barbara, Santa Barbara, California 93106, USA}

\author{Vladimir Iltchenko}
\affiliation{Jet Propulsion Laboratory, California Institute of Technology, Pasadena, California, USA}

\author{Owen Miller}
\affiliation{Department of Applied Physics, Yale University, New Haven, Connecticut 06520, USA}

\author{Andrey Matsko}
\affiliation{Jet Propulsion Laboratory, California Institute of Technology, Pasadena, California, USA}

\author{Steven Bowers}
\affiliation{Department of Electrical and Computer Engineering, University of Virginia, Charlottesville, VA 22904, USA}

\author{Peter T. Rakich}
\affiliation{Department of Applied Physics, Yale University, New Haven, Connecticut 06520, USA}

\author{Joe C. Campbell}
\affiliation{Department of Electrical and Computer Engineering, University of Virginia, Charlottesville, Virginia 22904, USA}

\author{John E. Bowers}
\affiliation{Department of Electrical and Computer Engineering, University of California, Santa Barbara, Santa Barbara, California 93106, USA}

\author{Kerry Vahala}
\affiliation{T. J. Watson Laboratory of Applied Physics, California Institute of Technology, Pasadena, CA 91125, USA}

\author{Franklyn Quinlan}
\affiliation{National Institute of Standards and Technology, 325 Broadway, Boulder, CO 80305, USA}

\author{Scott A. Diddams}
\affiliation{National Institute of Standards and Technology, 325 Broadway, Boulder, CO 80305, USA}
\affiliation{Department of Physics, University of Colorado Boulder, 440 UCB Boulder, CO 80309, USA}
\affiliation{Electrical Computer and Energy Engineering, University of Colorado, Boulder, CO 80309, USA}

\begin{abstract}

Numerous modern technologies are reliant on the low-phase noise and exquisite timing stability of microwave signals. Substantial progress has been made in the field of microwave photonics, whereby low noise microwave signals are generated by the down-conversion of ultra-stable optical references using a frequency comb. Such systems, however, are constructed with bulk or fiber optics and are difficult to further reduce in size and power consumption. Our work addresses this challenge by leveraging advances in integrated photonics to demonstrate low-noise microwave generation via two-point optical frequency division. Narrow linewidth self-injection locked integrated lasers are stabilized to a miniature Fabry-P\'{e}rot cavity, and the frequency gap between the lasers is divided with an efficient dark-soliton frequency comb. The stabilized output of the microcomb is photodetected to produce a microwave signal at 20 GHz with phase noise of -96 dBc/Hz at 100 Hz offset frequency that decreases to -135 dBc/Hz at 10 kHz offset--values which are unprecedented for an integrated photonic system. All photonic components can be heterogeneously integrated on a single chip, providing a significant advance for the application of photonics to high-precision navigation, communication and timing systems.
\end{abstract}

\maketitle

\section*{Introduction}
Low noise microwave signals with high timing stability are a critical enabler of modern science and multiple technologies of broad societal impact. Positioning and navigation, advanced communications, high-fidelity radar and sensing, and high-performance atomic clocks are all dependent upon low-phase noise microwave signals. These rapidly developing technologies are constantly intensifying the demand for microwave sources beyond current capabilities, while imposing harsher restrictions on system size and power consumption. In this landscape, photonic lightwave systems provide unique advantages over more conventional electronic approaches for generating low-noise microwaves. In particular, the extremely low loss and high quality factors of photonic resonators are fundamental to electromagnetic (EM) oscillators with the lowest noise and highest spectral purity~\cite{matei2017}. Coupled to this is the introduction and rapid development of frequency combs in the last few decades that enable seamless coherent synthesis across the full EM spectrum~\cite{diddams2020optical}. This includes the frequency division of a 200-500 THz optical carrier down to a 10 GHz microwave with unrivaled long- and short-term stability~\cite{nakamura2020coherent, xie2017photonic, fortier2011generation, kalubovilage2022x}.  

\begin{figure}[!htb]
\centering
\includegraphics[width=\linewidth]{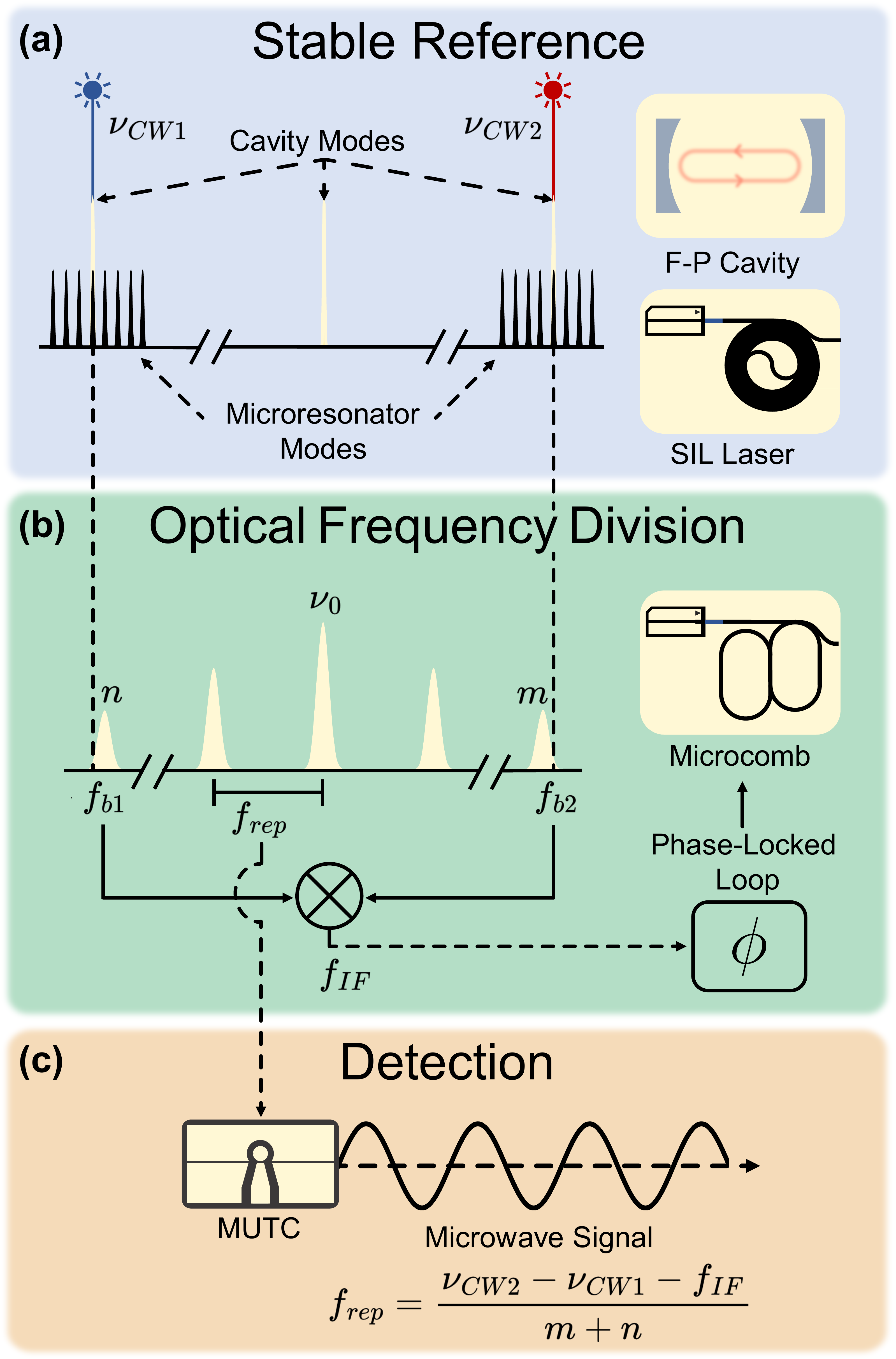}
\caption{\textbf{Concept of 2-point optical frequency division for low-noise microwave generation.} \textbf{(a)} Two semiconductor lasers are self-injection locked (SIL) to chip-based spiral resonators. The optical modes of the spiral resonators are aligned, using temperature control, to the modes of the high-finesse Fabry-P\'{e}rot (F-P) cavity for PDH locking. \textbf{(b)} A microcomb is generated in a coupled dual-ring resonator and is heterodyned with the two stabilized lasers. The beat-notes are mixed to produce an intermediate frequency, $f_{IF}$, that is phase locked by feedback to the current supply of the microcomb seed laser. \textbf{(c)} A modified uni-travelling carrier (MUTC) photodetector chip is used to convert the microcomb's optical output to a 20 GHz microwave signal.}
\label{fig:architecture}
\end{figure}

However, a significant challenge of these approaches is the relatively large size and power consumption that restrict their use to laboratory environments. Greater impact and widespread use can be realized with a low-noise microwave generator that has a compact and portable form factor for operation in remote and mobile platforms. Our work addresses and overcomes this challenge through the optimal implementation of 2-point optical frequency division (2P-OFD) with integrated photonic components. Figure~\ref{fig:architecture} illustrates the principles and advantages of the 2P-OFD system we introduce. Our work provides a means to significantly reduce microwave phase noise in a volume of 10's of milliliters instead of 10's of liters, while similarly reducing the required power by a factor $10^3$ to the 1 Watt level.  

All OFD systems start with a stable optical frequency reference. Typically this is a laboratory fiber or solid state laser that is frequency-stabilized to a large evacuated Fabry-P\'{e}rot (F-P) cavity~\cite{matei2017,Martin2012}. Instead, we introduce an optimal combination of low-noise chip-integrated semiconductor lasers~\cite{jin2021hertz, xiang2021high} and a new F-P concept that can be miniaturized to $<1 \rm{cm}^3$ and chip-integrated without the need for high-vacuum enclosure~\cite{jin2022micro, McLemore2022, kelleher2023compact, liu2023cmr}. The frequency noise of two semiconductor lasers near 1560 nm is reduced by 40 dB through self-injection locking (SIL) to high-Q Si$_3$N$_4$ spiral resonators~\cite{Li:21, guo2022chip}. This passive stabilization of the SIL laser enables further noise reduction, by up to 60 dB, through Pound–Drever–Hall (PDH) locking to a miniature F-P cavity, reaching the cavity’s thermal noise limit~\cite{drever1983laser, guo2022chip}. 

In OFD, the optical phase noise of the reference is then reduced by the square of the ratio of its frequency to that of the microwave output. This is a powerful means for noise reduction by a factor as large as $(2\times10^{14}/1\times10^{10})^2=4\times10^8$, or equivalently 86 dB. 
However, significant electrical power is required for generating a frequency comb spanning $\sim$200 THz with mode spacing of 10-20 GHz. Instead, we employ the simplified approach of 2P-OFD~\cite{swann2011microwave, papp2014microresonator, li2014electro, kwon2022ultrastable}, with comb bandwidth on the order of 1 THz. This results in a lower division factor, but also significantly reduced size and power requirements. Such a trade-off still allows us to reach an unprecedented microwave phase noise level with integrated photonics due to the intrinsic low noise of the optical references employed.  

In our system, the frequency division is implemented with another injection-locked laser that generates a microcomb in a zero group velocity dispersion (GVD) resonator, engineered by two coupled rings in a Vernier configuration~\cite{ji2023engineered}. 
The microcomb operates without the need for optical amplification, and $\sim$30\% of the input pump power of $\sim$100 mW is efficiently transferred to the comb which spans nearly 10 nm. 2P-OFD in then implemented by heterodyning the two SIL lasers with the closest comb teeth to produce two beat-notes. These are mixed to provide a servo control signal at an intermediate frequency which is independent of the microcomb center frequency. Upon phase locking of the intermediate frequency, the noise of the microcomb is dramatically reduced. Photodetection of the stabilized microcomb output with a high-power and high-linearity modified uni-traveling-carrier (MUTC) photodetector~\cite{zang2018reduction} provides a 20 GHz microwave signal with phase noise of -135 dBc/Hz at 10 kHz offset frequency. This level of noise has not been achieved previously for a system that employs integrated photonic components. We note that the critical photonic devices used in our system can be further integrated to a single chip without the need for fiber or semiconductor amplifiers or optical isolators, providing ultrastable microwave generation in a compact form-factor. This advance is important for future applications of high-performance microwave sources with compact size and low-power usage that will operate beyond the research labs.  

\begin{figure*}[!htb]
\centering
\includegraphics[width=\linewidth]{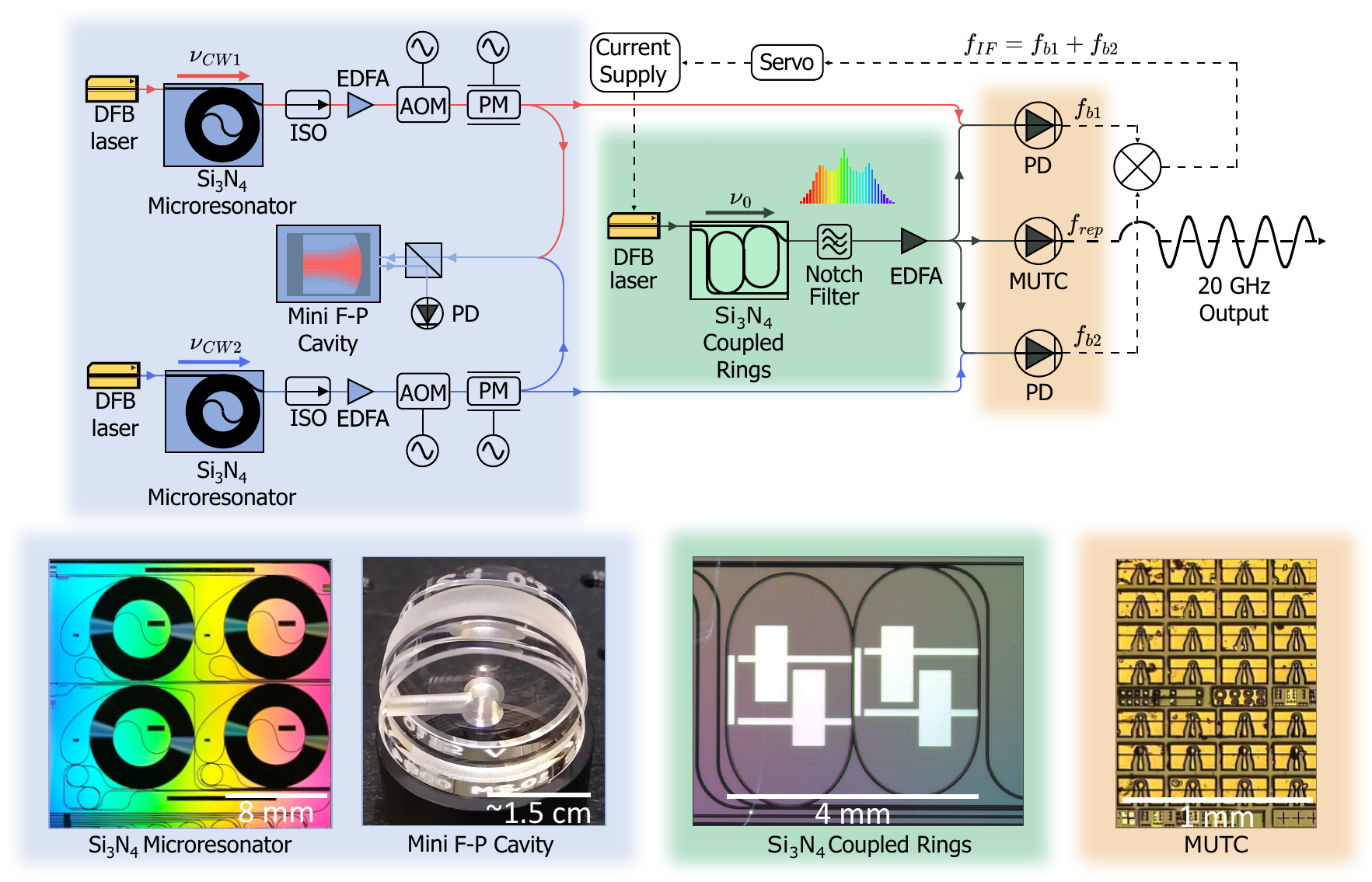}
\caption{\textbf{Experimental setup.} Two distributed feedback (DFB) lasers at 1557.3 nm and 1562.5 nm are self-injection locked to Si$_3$N$_4$ spiral resonators, amplified, and locked to the same miniature Fabry-P\'{e}rot cavity. A 6 nm broad frequency comb with a $\sim$20 GHz repetition rate is generated in a coupled-rings resonator. The microcomb is seeded by an integrated DFB laser, which is self-injection locked to the coupled-ring microresonator. The frequency comb passes through a notch filter to suppress the central line and then amplified to 60 mW total optical power.
 The frequency comb is split to beat with each of the PDH-locked SIL CW references. Two beat-notes are amplified, filtered and then mixed together to produce $f_{IF}$, which is phase locked to a reference frequency. The feedback for microcomb stabilization is provided to the current supply of the microcomb seed laser. Lastly, part of the generated microcomb is detected in a modified uni-travelling carrier (MUTC) detector to extract the low-noise 20 GHz signal.
\textbf{Bottom row:} Photographs of the key photonic components used in low noise microwave generation.
}
\label{fig2:exp_set}
\end{figure*}

\section*{Experiment and Results}
A technical illustration of the setup used for frequency comb stabilization and stable microwave generation is shown in Fig.~\ref{fig2:exp_set}. Here we elaborate on the operation and characteristics of the key components, concluding with a description of how they fit together cohesively to produce low phase noise microwave signals.
 
\textbf{Miniature Fabry-P\'{e}rot cavity:} The phase and frequency stability of the generated microwave signal is ultimately derived from that of the ultrastable optical reference. The lowest noise optical references are lasers locked to vacuum-gap F-P cavities, where fractional frequency stability as low as 4$\times$ 10$^{-17}$ has been demonstrated with 212 mm long, cryogenic cavity systems~\cite{matei2017}. Instead, we employ an integrable cavity design based on a compact, rigidly held cylindrical F-P optical reference cavity that supports fractional frequency stability at the 10$^{-14}$ level~\cite{kelleher2023compact}. Ultra-low expansion glass with 1 m radius of curvature and an ultra-low expansion glass spacer compose the 6.3 mm long cavity with finesse of $\sim$900,000 (Q$\sim$5 billion) and overall volume of less than 9 cm$^3$. The cavity is thermal noise limited for offset frequencies ranging from 1 Hz to 10 kHz. Moreover, the relative phase noise between two reference lasers locked to the same cavity takes advantage of large common-mode rejection (CMR), reaching 40 dB rejection for cavity modes spaced by 1 THz~\cite{liu2023cmr}. When combined with 2P-OFD, the cavity noise is expected to be reduced by $\sim$80 dB when projected onto our microwave carrier. 

\begin{figure*}[t]
\centering
\includegraphics[width=\linewidth]{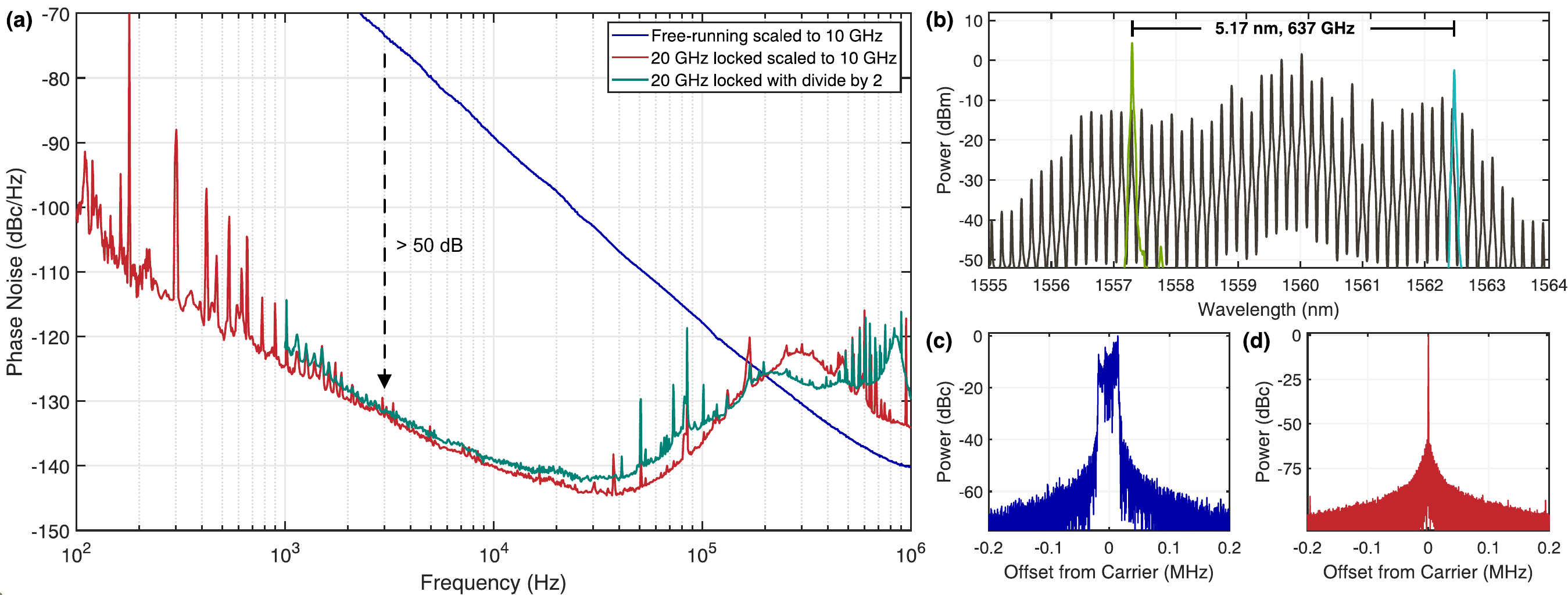}
\caption{\textbf{Microcomb characterisation.} \textbf{(a)} Single side-band (SSB) Phase noise scaled to 10 GHz of free-running 20 GHz microcomb (blue), locked 20 GHz microwave (red), and locked 20 GHz microwave after regenerative frequency division by 2 (green).  
\textbf{(b)} Optical spectrum of microcomb (grey), and SIL lasers (green and turquoise). RF spectra of 20 GHz signal \textbf{(c)} free-running (RBW 100 Hz) and \textbf{(d)} locked (RBW 1 Hz).
}
\label{fig3:phase_noise}
\end{figure*}

\textbf{Self-injection locked lasers:} To achieve high stability performance in the proposed system, it is crucial to use narrow-linewidth and frequency-stable lasers. This is because electrical noise from the individual laser locking circuits does not experience CMR, and this noise is reduced only by the 2P-OFD. To address this issue and reach the thermal noise floor of the F-P cavity, we employ SIL lasers that are both integrable and have noise performance equivalent to much larger laboratory fiber lasers\cite{jin2021hertz,Li:21,guo2022chip}. With reference to Fig.~\ref{fig2:exp_set}, two semiconductor DFBs are pre-stabilized by self-injection locking to a high-Q Si$_3$N$_4$ spiral resonator~\cite{Li:21}. When the forward and backward fields between the DFB lasers and the Si$_3$N$_4$ resonators are in phase, resonant backscattered light is then fed back to the DFB, anchoring each laser wavelength to the corresponding Si$_3$N$_4$ resonance and significantly suppressing its frequency noise. The length and Q of the integrated resonator determines the ultimate phase noise of the SIL laser~\cite{li1989analysis}, and here we employ a spiral with length of 1.41 m that is fabricated on $\sim 1 {\rm cm^2}$ of silicon. The free-spectral range (FSR) is 135 MHz. The intrinsic and loaded Q factors of the two spiral resonators are 164 and 126 million, respectively, enabling phase noise of -85~dBc/Hz at 10 kHz offset~\cite{Li:21}. The output of the SIL lasers are amplified using commercial EDFAs to $\sim$30 mW and then further stabilized via PDH locking to the miniature F-P cavity. In this setup, the primary actuator for PDH stabilization is an acousto-optic modulator; however, the PDH error signal is also fed back to the electro-optical modulator (EOM) to further increase the bandwidth and noise reduction of the PDH servo~\cite{endo2018residual}. The in-loop phase noise of the PDH locking of SIL lasers are presented in Supplementary Information.

\textbf{Microcomb:} Robust and low-noise optical frequency comb generation with 10-20 GHz repetition rate and broad optical coverage is challenging. Here, we use a Si$_3$N$_4$ microresonator fabricated at a CMOS foundry to generate mode-locked microcombs~\cite{jin2021hertz}. 
To produce dark soliton microcombs with higher bandwidth, we use a dual coupled-ring resonator with FSR of 20~GHz~\cite{ji2023engineered}, where the zero GVD wavelength is tuned to $\sim$1560 nm using integrated heaters~\cite{ji2023integrated}. In addition, such microcomb states have high pump-to-comb conversion efficiency, benefiting microwave generation in a low-SWaP system. To generate the comb, a commercial DFB laser without optical amplification is self-injection locked to the dual coupled-ring resonator, which narrows the linewidth of the pump laser and generates a reasonably stable 20 GHz comb~\cite{pavlov2018narrow,shen2020integrated,jin2021hertz}. Following the coupled rings, a notch filter is used to suppress the central (seed) comb line in order to avoid saturation in an EDFA, which amplifies the frequency comb up to 60 mW (Fig.~\ref{fig2:exp_set}). A typical comb spectrum after the EDFA is shown in Fig.~\ref{fig3:phase_noise}(b).  

\begin{figure}[!tb]
\centering
\includegraphics[width=\linewidth]{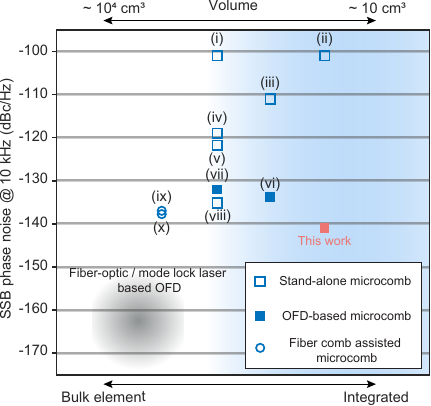}
\caption{\textbf{Phase noise comparison of microwave generations based on microcomb, scaled to 10 GHz.} 
The platforms are categorized based on the integration capability of microcomb generator and the reference laser source, excluding the inter-connecting optical/electrical parts. Filled (blank) squares are based on the OFD (stand-alone microcomb) approach. 
(i) 22 GHz on-chip silica microcomb \cite{yi2015soliton}, 
(ii) 5 GHz hybridly integrated Si$_3$N$_4$ microcomb~\cite{jin2021hertz}, 
(iii) 10.8 GHz photonic chip-based Si$_3$N$_4$ microcomb~\cite{liu2020photonic},
(iv) 22 GHz on-chip silica microcomb~\cite{li2012low}, 
(v) miniaturized MgF$_2$ microcomb~\cite{liang2015high}, 
(vi) 100 GHz photonic chip-based Si$_3$N$_4$ microcomb~\cite{sun2023integrated},
(vii) 22~GHz Fibre-stabilised SiO$_2$ microcomb~\cite{kwon2022ultrastable},
(viii) miniaturized MgF$_2$ microcomb~\cite{matsko2016turn},
(ix) 14 GHz MgF$_2$ microcomb pumped by an ultrastable laser \cite{weng2019spectral},  
(x) 14 GHz microcomb-based transfer oscillator~\cite{lucas2020ultralow}.  .
}
\label{fig4:pn_comparison}
\end{figure}

\textbf{Microcomb stabilization and microwave generation:}
As outlined above, in this work we use 2-point locking to realize optical frequency division for phase noise reduction. With appropriate fiber-optic couplers and filters, we design a receiver system to separate heterodyne beat-notes and 20 GHz microwave generation. The two beat-notes between the microcomb and each CW laser are given by $f_{\rm b1} = \nu_0 - n\cdot f_{\rm rep} - \nu_{\rm CW1} $ and $f_{\rm b2} = \nu_{\rm CW2} - \nu_0 - m\cdot f_{\rm rep}$ (see Fig.~\ref{fig:architecture}). These beats are filtered, amplified, and mixed together to produce the intermediate frequency ($f_{\rm IF}$), which is then phase-locked to a stable microwave reference, via feedback to the current of the microcomb seed laser~\cite{ji2023engineered}. The stabilization of $f_{\rm IF}$ is the final step to generate a low-noise microwave via 2P-OFD:
\begin{equation*}
    f_{\rm IF} = f_{\rm b1} + f_{\rm b2} = \nu_{\rm CW2} - \nu_{\rm CW1} + (n+m)f_{\rm rep}
\end{equation*}
in which $\nu_{CW1}$ and $\nu_{CW1}$ are the carrier frequencies of the optical references, and $n+m$ is the value of the optical frequency division (32 in our case), which amounts to 30 dB of noise reduction. Note that 2-point locking does not depend on the noise of the microcomb central frequency $\nu_0$. Thus, the instability of the microcomb repetition rate can be represented as:
\begin{equation*}
    \delta f_{\rm rep}^2 = \frac{\delta (\nu_{\rm CW2} - \nu_{\rm CW1})^2 + \delta f_{\rm IF}^2}{(n+m)^2}. 
\end{equation*}
While the phase noise of $f_{IF}$, $\nu_{\rm CW2}$ and $\nu_{\rm CW2}$ are all reduced by 2P-OFD, their servo control and residual noise can be limiting factors in the achievable microwave phase noise. These, and other limitations to the achievable phase, noise are discussed in greater detail in the Supplementary Information. 
 
The stabilized microcomb output is directed to a MUTC photodiode, which provides exceptional linearity and large microwave powers~\cite{peng2020photonic, Xie:14}. We tune the bias voltage of the MUTC operation for $\approx40$ dB  rejection of amplitude-to-phase noise conversion, while generating a 20 GHz power of -10 dBm at 5 mA of average photocurrent. The 20~GHz microwave signal is filtered and amplified to +3 dBm and sent to a measurement system where the 20~GHz signal is first downconverted to 47 MHz by mixing with an ultra-stable microwave from a self-referenced Er:fiber mode-locked laser~\cite{nakamura2020coherent}. The 47 MHz signal is characterized with a commercial phase noise analyzer (53100A Microchip) and results are presented in Fig.~\ref{fig3:phase_noise}(a). Here we have scaled the measured 20 GHz phase noise to 10 GHz by subtracting 6 dB. This yields -102 dBc/Hz at 100 Hz, which decreases to -141 dBc/Hz at 10 kHz and reaches its minimum of -145 dBc/Hz at 40 kHz frequency offset. We also compare with a 10 GHz carrier that we generate from 20 GHz with a regenerative divide-by-2 circuit. Compared to the free-running microcomb generation, we achieved more than 50 dB phase noise improvement for offset frequencies below 10 kHz. Additional details on the phase noise measurement and divider are in the Supplementary Information.

\section*{Discussion and further integration} 

\begin{figure*}[!htb]
\centering
\includegraphics[width=0.95\linewidth]{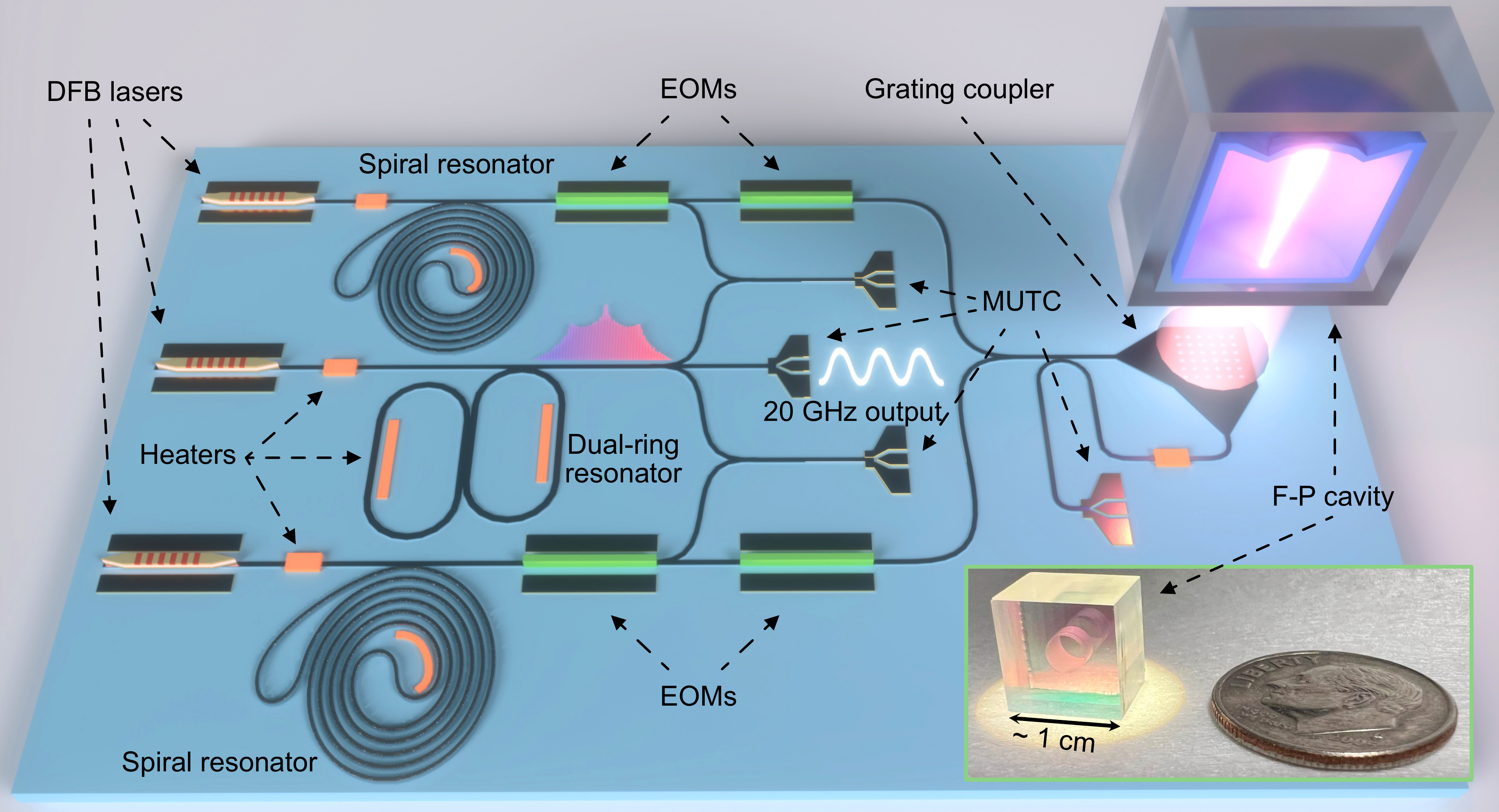}
\caption{\textbf{Schematic design of a photonic microwave oscillator on a single chip.} The integrated system employs the same key photonic elements used in this work. Two spiral-resonator SIL lasers are PDH-locked to the same micro F-P cavity with two EOMs in series for each SIL laser---the first for fast phase correction and the second for PDH sidebands. The right side of the schematic shows the F-P cavity interface, where the two SIL laser paths are fed through an interferometer with an embedded polarization splitting grating. This serves as a reflection cancellation circuit while also shaping the planar waveguide mode to match the F-P mode~\cite{cheng2022chip}. The reflection from the F-P cavity is then detected by the right-most detector. Inset: Photo of the miniature F-P cavity, consisted of micro-fabricated mirrors~\cite{jin2022micro}, with overall volume of $\sim$1 cm$^3$.
}
\label{fig5:integration}
\end{figure*}

Figure~\ref{fig4:pn_comparison} places the level of phase noise we achieve in context with other photonic approaches, including recent works based on microcombs and mode-locked laser frequency combs. The comparison is classified by level of photonic integration of the microcomb source and pumping/reference lasers, as applicable. It is also noted that some of the microcomb systems require the assistance of a fiber-based frequency comb (Fig.~\ref{fig4:pn_comparison} (ix), (x)~\cite{weng2019spectral,lucas2020ultralow}). The phase noise performance of other systems, which could be chip integrated (Fig.~\ref{fig4:pn_comparison} (ii), (iii)~\cite{jin2021hertz,liu2020photonic}), is more than 30 dB greater than the results we present, with the exception of the recent work by Sun, {\it et al.} (vi)~\cite{sun2023integrated}. Other notable works on low-noise microwave generation in a low-SWaP systems, which are not shown in Fig.~\ref{fig4:pn_comparison}, include 'quite point' operation~\cite{yi2017single,lucas2020ultralow,yang2021dispersive,yao2022soliton}, single laser optical frequency division~\cite{zhao2023all}, and commercial products~\cite{OEwaves, li2023small, Quantx}. 

To the best of our knowledge, this work provides the best phase noise performance in the frequency range of 200 Hz - 40 kHz for microcomb-based systems. Importantly, it does so with integrated photonic components that can all be further combined onto a single chip with total volume of the photonic components of approximately $\rm{1 cm^3}$. A concept of such a fully-integrated system is shown in Fig.~\ref{fig5:integration}(a) and would consist of heterogeneously integrated lasers at 1560 nm~\cite{guo2022chip}, spiral resonators~\cite{Li:21} for SIL, a coupled-ring microcomb resonator~\cite{ji2023engineered}, photodetectors~\cite{zang2018reduction} and a micro-fabricated F-P cavity that does not require high vacuum~\cite{jin2022micro,liu2023cmr}. 

Previous work already lays out the steps for heterogeneous integration of lasers, resonators and photodiodes. For example, InP lasers and Si$_3$N$_4$ resonators have been integrated on the same chip with coupling between the optical gain and low-loss waveguide layers facilitated by adiabatic tapers, reaching down to 0.5~dB/m with a second deeply buried Si$_3$N$_4$ waveguide layer~\cite{xiang2021high,xiang2023three}. This same  heterogeneously integration promises isolator-free operation~\cite{xiang2023three}. A similar strategy has been employed for laser integration with thick Si$_3$N, anomalous dispersion microcombs on the same chip~\cite{xiang2021laser}, which can be applied to the zero-dispersion microcombs used in this work. Laser integration with modulators, detectors, and optical amplifiers based on the heterogeneous InP/Si platform has also been previously demonstrated~\cite{xie2019heterogeneous, davenport2016heterogeneous} and can be utilized for full integration of the optical components comprising the PDH locking system~\cite{idjadi2017integrated}. Note that previously, the various laser, modulator, and detector InP epitaxial stacks were modified so that they could be etched simultaneously; to account for the MUTC photodiodes~\cite{zang2018reduction}, similar modifications will need to be done to ensure process compatibility. 

The integration of the active and passive components on a single platform greatly reduces loss (between fiber and chip) and removes the need for optical amplifiers we have employed in the present work. In such a case, a few tens of mW of optical power is required to pump the resonator such that a comb with a few mW of optical power and several $\mu$W per mode is realistic. Additionally, for the SIL lasers only several mW of DFB optical power is required to provide a hundred $\mu$W of optical power to heterodyne with the comb and achieve the SNR necessary to match the performance presented in this work. For recent integrated lasers, these powers are realistic ~\cite{xiang2021high}. Additional considerations on required optical power are discussed in Supplementary Information.

Integration of the FP cavity has been an outstanding challenge, but recent developments in micro-fabricated mirrors~\cite{jin2022micro} and compact thermal-noise limited F-P designs~\cite{McLemore2022} provide new integration opportunities. Critically, it has been shown that 2P-OFD does not require FP operation in high-vacuum due to common mode rejection~\cite{liu2023cmr}, significantly simplifying future integration. 
Fig.~\ref{fig5:integration} shows a $\rm{1 cm^3}$ cavity with fabricated micro-mirrors and details on an integration strategy with the the SIL lasers and microcomb. A planar waveguide feeds an inverse-designed polarization splitting grating embedded in an interferometer, which serves to shape the beam for coupling light to the cavity while also providing the cavity-reflected PDH locking signal and laser isolation~\cite{cheng2022chip}. Preliminary measurements described in the Supplement demonstrate the feasibility of this approach. The F-P cavity and a GRIN lens can be bonded on top of the polarization splitting grating in a hybrid flip-chip fashion for a single-chip, cavity-integrated low-noise microwave generator unit. 

In this integration scheme, the AOMs can be replaced with a combniation of slow feedback to the integrated heaters in the spirals and fast feedback to an external EOM. The thermal tuning can reach bandwidth of a few kHz~\cite{joshi2016thermally, xiang2023three}, while the fast feedback with nearly 1 MHz of bandwidth could be provided by the EOM~\cite{endo2018residual}. We estimate that this combination can provide 40 dB feedback gain at 10 kHz offset frequency to match the phase noise presented in this work. 

In summary, we have demonstrated an integrated photonic approach to OFD that produces 20 GHz microwave signals with phase noise of -135 dBc/Hz at 10 kHz offset. This is a value typical of much larger existing commercial systems. This is accomplished with a unique combination of low-noise integrated lasers, an efficient dark-soliton frequency comb and new advances in a miniature F-P optical cavity. Significantly, our approach provides a route to full integration on single chip with volume of the photonic components on the order of ${\rm 1 cm^3}$. This advance in integrated photonic low noise microwave generation holds promise for compact, portable and low-cost microwave synthesis for a wide variety of demanding applications in navigation, communications and precise timing.

\bibliographystyle{ieeetr}
\bibliography{Gryphon_paper}

\begin{thebibliography}{10}

\bibitem{matei2017}
D.~Matei, T.~Legero, S.~H{\"a}fner, C.~Grebing, R.~Weyrich, W.~Zhang,
  L.~Sonderhouse, J.~Robinson, J.~Ye, F.~Riehle, {\em et~al.}, ``1.5 $\mu$m
  lasers with sub-10 mhz linewidth,'' {\em Physical Review Letters}, vol.~118,
  no.~26, p.~263202, 2017.

\bibitem{diddams2020optical}
S.~A. Diddams, K.~Vahala, and T.~Udem, ``Optical frequency combs: Coherently
  uniting the electromagnetic spectrum,'' {\em Science}, vol.~369, no.~6501,
  p.~eaay3676, 2020.

\bibitem{nakamura2020coherent}
T.~Nakamura, J.~Davila-Rodriguez, H.~Leopardi, J.~A. Sherman, T.~M. Fortier,
  X.~Xie, J.~C. Campbell, W.~F. McGrew, X.~Zhang, Y.~S. Hassan, D.~Nicolodi,
  K.~Beloy, A.~D. Ludlow, S.~A. Diddams, and F.~Quinlan, ``Coherent optical
  clock down-conversion for microwave frequencies with $10^{-18}$
  instability,'' {\em Science}, vol.~368, no.~6493, pp.~889--892, 2020.

\bibitem{xie2017photonic}
X.~Xie, R.~Bouchand, D.~Nicolodi, M.~Giunta, W.~H{\"a}nsel, M.~Lezius,
  A.~Joshi, S.~Datta, C.~Alexandre, M.~Lours, {\em et~al.}, ``Photonic
  microwave signals with zeptosecond-level absolute timing noise,'' {\em Nature
  Photonics}, vol.~11, no.~1, pp.~44--47, 2017.

\bibitem{fortier2011generation}
T.~M. Fortier, M.~S. Kirchner, F.~Quinlan, J.~Taylor, J.~Bergquist,
  T.~Rosenband, N.~Lemke, A.~Ludlow, Y.~Jiang, C.~Oates, and S.~A. Diddams,
  ``Generation of ultrastable microwaves via optical frequency division,'' {\em
  Nature Photonics}, vol.~5, no.~7, pp.~425--429, 2011.

\bibitem{kalubovilage2022x}
M.~Kalubovilage, M.~Endo, and T.~R. Schibli, ``X-band photonic microwaves with
  phase noise below- 180 dbc/hz using a free-running monolithic comb,'' {\em
  Optics Express}, vol.~30, no.~7, pp.~11266--11274, 2022.

\bibitem{Martin2012}
M.~J. Martin and J.~Ye, ``High-precision laser stabilization via optical
  cavities,'' {\em Optical Coatings and Thermal Noise in Precision Measurement,
  GM Harry, T. Bodiya, R. DeSalvo., Eds}, pp.~237--258, 2012.

\bibitem{jin2021hertz}
W.~Jin, Q.-F. Yang, L.~Chang, B.~Shen, H.~Wang, M.~A. Leal, L.~Wu, M.~Gao,
  A.~Feshali, M.~Paniccia, K.~J. Vahala, and J.~E. Bowers, ``Hertz-linewidth
  semiconductor lasers using cmos-ready ultra-high-q microresonators,'' {\em
  Nature Photonics}, vol.~15, no.~5, pp.~346--353, 2021.

\bibitem{xiang2021high}
C.~Xiang, J.~Guo, W.~Jin, L.~Wu, J.~Peters, W.~Xie, L.~Chang, B.~Shen, H.~Wang,
  Q.-F. Yang, D.~Kinghorn, M.~Paniccia, K.~J. Vahala, P.~A. Morton, and J.~E.
  Bowers, ``High-performance lasers for fully integrated silicon nitride
  photonics,'' {\em Nature Communications}, vol.~12, no.~1, pp.~1--8, 2021.

\bibitem{jin2022micro}
N.~Jin, C.~A. McLemore, D.~Mason, J.~P. Hendrie, Y.~Luo, M.~L. Kelleher,
  P.~Kharel, F.~Quinlan, S.~A. Diddams, and P.~T. Rakich, ``Micro-fabricated
  mirrors with finesse exceeding one million,'' {\em Optica}, vol.~9, no.~9,
  pp.~965--970, 2022.

\bibitem{McLemore2022}
C.~A. McLemore, N.~Jin, M.~L. Kelleher, J.~P. Hendrie, D.~Mason, Y.~Luo,
  D.~Lee, P.~Rakich, S.~A. Diddams, and F.~Quinlan, ``Miniaturizing ultrastable
  electromagnetic oscillators: Sub-$10^{14}$ frequency instability from a
  centimeter-scale fabry-perot cavity,'' {\em Physical Review Applied},
  vol.~18, p.~054054, 11 2022.

\bibitem{kelleher2023compact}
M.~L. Kelleher, C.~A. McLemore, D.~Lee, J.~Davila-Rodriguez, S.~A. Diddams, and
  F.~Quinlan, ``Compact, portable, thermal-noise-limited optical cavity with
  low acceleration sensitivity,'' {\em Optics Express}, vol.~31, no.~7,
  pp.~11954--11965, 2023.

\bibitem{liu2023cmr}
Y.~Liu, C.~A. McLemore, M.~Kelleher, D.~Lee, T.~Nakamura, N.~Jin, S.~Schima,
  P.~Rakich, S.~A. Diddams, and F.~Quinlan, ``Thermal-noise-limited, compact
  optical reference cavity operated without a vacuum enclosure,'' {\em arXiv
  preprint arXiv:2307.04758}, 2023.

\bibitem{Li:21}
B.~Li, W.~Jin, L.~Wu, L.~Chang, H.~Wang, B.~Shen, Z.~Yuan, A.~Feshali,
  M.~Paniccia, K.~J. Vahala, and J.~E. Bowers, ``Reaching fiber-laser coherence
  in integrated photonics,'' {\em Opt. Lett.}, vol.~46, pp.~5201--5204, Oct
  2021.

\bibitem{guo2022chip}
J.~Guo, C.~A. McLemore, C.~Xiang, D.~Lee, L.~Wu, W.~Jin, M.~Kelleher, N.~Jin,
  D.~Mason, L.~Chang, {\em et~al.}, ``Chip-based laser with 1-hertz integrated
  linewidth,'' {\em Science Advances}, vol.~8, no.~43, p.~eabp9006, 2022.

\bibitem{drever1983laser}
R.~W. Drever, J.~L. Hall, F.~V. Kowalski, J.~Hough, G.~Ford, A.~Munley, and
  H.~Ward, ``Laser phase and frequency stabilization using an optical
  resonator,'' {\em Applied Physics B}, vol.~31, pp.~97--105, 1983.

\bibitem{swann2011microwave}
W.~C. Swann, E.~Baumann, F.~R. Giorgetta, and N.~R. Newbury, ``Microwave
  generation with low residual phase noise from a femtosecond fiber laser with
  an intracavity electro-optic modulator,'' {\em Optics Express}, vol.~19,
  no.~24, pp.~24387--24395, 2011.

\bibitem{papp2014microresonator}
S.~B. Papp, K.~Beha, P.~Del’Haye, F.~Quinlan, H.~Lee, K.~J. Vahala, and S.~A.
  Diddams, ``Microresonator frequency comb optical clock,'' {\em Optica},
  vol.~1, no.~1, pp.~10--14, 2014.

\bibitem{li2014electro}
J.~Li, X.~Yi, H.~Lee, S.~A. Diddams, and K.~J. Vahala, ``Electro-optical
  frequency division and stable microwave synthesis,'' {\em Science}, vol.~345,
  no.~6194, pp.~309--313, 2014.

\bibitem{kwon2022ultrastable}
D.~Kwon, D.~Jeong, I.~Jeon, H.~Lee, and J.~Kim, ``Ultrastable microwave and
  soliton-pulse generation from fibre-photonic-stabilized microcombs,'' {\em
  Nature Communications}, vol.~13, no.~1, p.~381, 2022.

\bibitem{ji2023engineered}
Q.-X. Ji, W.~Jin, L.~Wu, Y.~Yu, Z.~Yuan, W.~Zhang, M.~Gao, B.~Li, H.~Wang,
  C.~Xiang, {\em et~al.}, ``Engineered zero-dispersion microcombs using
  cmos-ready photonics,'' {\em Optica}, vol.~10, no.~2, pp.~279--285, 2023.

\bibitem{zang2018reduction}
J.~Zang, X.~Xie, Q.~Yu, Z.~Yang, A.~Beling, and J.~C. Campbell, ``Reduction of
  amplitude-to-phase conversion in charge-compensated modified unitraveling
  carrier photodiodes,'' {\em Journal of Lightwave Technology}, vol.~36,
  no.~22, pp.~5218--5223, 2018.

\bibitem{li1989analysis}
H.~Li and N.~Abraham, ``Analysis of the noise spectra of a laser diode with
  optical feedback from a high-finesse resonator,'' {\em IEEE Journal of
  Quantum Electronics}, vol.~25, no.~8, pp.~1782--1793, 1989.

\bibitem{endo2018residual}
M.~Endo and T.~R. Schibli, ``Residual phase noise suppression for
  pound-drever-hall cavity stabilization with an electro-optic modulator,''
  {\em OSA Continuum}, vol.~1, no.~1, pp.~116--123, 2018.

\bibitem{ji2023integrated}
Q.~Ji, W.~Jin, J.~Guo, L.~Wu, P.~Liu, A.~Feshali, M.~Paniccia, J.~Bowers, and
  K.~Vahala, ``Integrated microcomb with broadband tunable normal and anomalous
  dispersion,'' in {\em Optica Nonlinear Optics Topical Meeting}, 2023.

\bibitem{pavlov2018narrow}
N.~Pavlov, S.~Koptyaev, G.~Lihachev, A.~Voloshin, A.~Gorodnitskiy, M.~Ryabko,
  S.~Polonsky, and M.~Gorodetsky, ``Narrow-linewidth lasing and soliton kerr
  microcombs with ordinary laser diodes,'' {\em Nature Photonics}, vol.~12,
  no.~11, pp.~694--698, 2018.

\bibitem{shen2020integrated}
B.~Shen, L.~Chang, J.~Liu, H.~Wang, Q.-F. Yang, C.~Xiang, R.~N. Wang, J.~He,
  T.~Liu, W.~Xie, J.~Guo, D.~Kinghorn, L.~Wu, Q.-X. Ji, T.~J. Kippenberg,
  K.~Vahala, and J.~E. Bowers, ``Integrated turnkey soliton microcombs,'' {\em
  Nature}, vol.~582, no.~7812, pp.~365--369, 2020.

\bibitem{yi2015soliton}
X.~Yi, Q.-F. Yang, K.~Y. Yang, M.-G. Suh, and K.~Vahala, ``Soliton frequency
  comb at microwave rates in a high-q silica microresonator,'' {\em Optica},
  vol.~2, no.~12, pp.~1078--1085, 2015.

\bibitem{liu2020photonic}
J.~Liu, E.~Lucas, A.~S. Raja, J.~He, J.~Riemensberger, R.~N. Wang, M.~Karpov,
  H.~Guo, R.~Bouchand, and T.~J. Kippenberg, ``Photonic microwave generation in
  the x-and k-band using integrated soliton microcombs,'' {\em Nature
  Photonics}, vol.~14, no.~8, pp.~486--491, 2020.

\bibitem{li2012low}
J.~Li, H.~Lee, T.~Chen, and K.~J. Vahala, ``Low-pump-power, low-phase-noise,
  and microwave to millimeter-wave repetition rate operation in microcombs,''
  {\em Physical Review Letters}, vol.~109, no.~23, p.~233901, 2012.

\bibitem{liang2015high}
W.~Liang, D.~Eliyahu, V.~S. Ilchenko, A.~A. Savchenkov, A.~B. Matsko,
  D.~Seidel, and L.~Maleki, ``High spectral purity kerr frequency comb radio
  frequency photonic oscillator,'' {\em Nature communications}, vol.~6, no.~1,
  p.~7957, 2015.

\bibitem{sun2023integrated}
S.~Sun, B.~Wang, K.~Liu, M.~Harrington, F.~Tabatabaei, R.~Liu, J.~Wang,
  S.~Hanifi, J.~S. Morgan, M.~Jahanbozorgi, {\em et~al.}, ``Integrated optical
  frequency division for stable microwave and mmwave generation,'' {\em arXiv
  preprint arXiv:2305.13575}, 2023.

\bibitem{matsko2016turn}
A.~Matsko, A.~Savchenkov, D.~Eliyahu, W.~Liang, E.~Dale, V.~Ilchenko, and
  L.~Maleki, ``Turn-key operation and stabilization of kerr frequency combs,''
  in {\em 2016 IEEE International Frequency Control Symposium (IFCS)},
  pp.~1--5, IEEE, 2016.

\bibitem{weng2019spectral}
W.~Weng, E.~Lucas, G.~Lihachev, V.~E. Lobanov, H.~Guo, M.~L. Gorodetsky, and
  T.~J. Kippenberg, ``Spectral purification of microwave signals with
  disciplined dissipative kerr solitons,'' {\em Physical Review Letters},
  vol.~122, no.~1, p.~013902, 2019.

\bibitem{lucas2020ultralow}
E.~Lucas, P.~Brochard, R.~Bouchand, S.~Schilt, T.~S{\"u}dmeyer, and T.~J.
  Kippenberg, ``Ultralow-noise photonic microwave synthesis using a soliton
  microcomb-based transfer oscillator,'' {\em Nature Communications}, vol.~11,
  no.~1, p.~374, 2020.

\bibitem{peng2020photonic}
Y.~Peng, K.~Sun, Y.~Shen, A.~Beling, and J.~C. Campbell, ``Photonic generation
  of pulsed microwave signals in the x-, ku-and k-band,'' {\em Optics Express},
  vol.~28, no.~19, pp.~28563--28572, 2020.

\bibitem{Xie:14}
X.~Xie, Q.~Zhou, K.~Li, Y.~Shen, Q.~Li, Z.~Yang, A.~Beling, and J.~C. Campbell,
  ``Improved power conversion efficiency in high-performance photodiodes by
  flip-chip bonding on diamond,'' {\em Optica}, vol.~1, pp.~429--435, Dec 2014.

\bibitem{cheng2022chip}
H.~Cheng, Z.~Dai, Y.~Zhou, F.~Ruesink, N.~Jin, D.~Mason, O.~Miller, and
  P.~Rakich, ``On-chip poor man’s isolator,'' in {\em Frontiers in Optics},
  pp.~FM5D--1, Optica Publishing Group, 2022.

\bibitem{yi2017single}
X.~Yi, Q.-F. Yang, X.~Zhang, K.~Y. Yang, X.~Li, and K.~Vahala, ``Single-mode
  dispersive waves and soliton microcomb dynamics,'' {\em Nature
  communications}, vol.~8, no.~1, p.~14869, 2017.

\bibitem{yang2021dispersive}
Q.-F. Yang, Q.-X. Ji, L.~Wu, B.~Shen, H.~Wang, C.~Bao, Z.~Yuan, and K.~Vahala,
  ``Dispersive-wave induced noise limits in miniature soliton microwave
  sources,'' {\em Nature communications}, vol.~12, no.~1, p.~1442, 2021.

\bibitem{yao2022soliton}
L.~Yao, P.~Liu, H.-J. Chen, Q.~Gong, Q.-F. Yang, and Y.-F. Xiao, ``Soliton
  microwave oscillators using oversized billion q optical microresonators,''
  {\em Optica}, vol.~9, no.~5, pp.~561--564, 2022.

\bibitem{zhao2023all}
Y.~Zhao, J.~K. Jang, K.~J. McNulty, X.~Ji, Y.~Okawachi, M.~Lipson, and A.~L.
  Gaeta, ``All-optical frequency division on-chip using a single laser,'' {\em
  arXiv preprint arXiv:2303.02805}, 2023.

\bibitem{OEwaves}
``Hi-\uppercase{Q X}-band \uppercase{OEO}.''
  \url{https://www.oewaves.com/oe3700}.
\newblock Accessed: 2023-06-18.

\bibitem{li2023small}
J.~Li and K.~Vahala, ``Small-sized, ultra-low phase noise photonic microwave
  oscillators at x-ka bands,'' {\em Optica}, vol.~10, no.~1, pp.~33--34, 2023.

\bibitem{Quantx}
``Ultra-low-noise microwave oscillator.''
  \url{https://www.quantxlabs.com/capabilities/product-development/ultra-low-phase-noise-oscillators/}.
\newblock Accessed: 2023-07-03.

\bibitem{xiang2023three}
C.~Xiang, W.~Jin, O.~Terra, B.~Dong, H.~Wang, L.~Wu, J.~Guo, T.~J. Morin,
  E.~Hughes, J.~Peters, {\em et~al.}, ``Three-dimensional integration enables
  ultra-low-noise, isolator-free si photonics,'' {\em arXiv preprint
  arXiv:2301.09989}, 2023.

\bibitem{xiang2021laser}
C.~Xiang, J.~Liu, J.~Guo, L.~Chang, R.~N. Wang, W.~Weng, J.~Peters, W.~Xie,
  Z.~Zhang, J.~Riemensberger, J.~Selvidge, T.~J. Kippenberg, and J.~E. Bowers,
  ``Laser soliton microcombs heterogeneously integrated on silicon,'' {\em
  Science}, vol.~373, no.~6550, pp.~99--103, 2021.

\bibitem{xie2019heterogeneous}
W.~Xie, T.~Komljenovic, J.~Huang, M.~Tran, M.~Davenport, A.~Torres, P.~Pintus,
  and J.~Bowers, ``Heterogeneous silicon photonics sensing for autonomous
  cars,'' {\em Optics express}, vol.~27, no.~3, pp.~3642--3663, 2019.

\bibitem{davenport2016heterogeneous}
M.~L. Davenport, S.~Skend{\v{z}}i{\'c}, N.~Volet, J.~C. Hulme, M.~J. Heck, and
  J.~E. Bowers, ``Heterogeneous silicon/iii--v semiconductor optical
  amplifiers,'' {\em IEEE Journal of Selected Topics in Quantum Electronics},
  vol.~22, no.~6, pp.~78--88, 2016.

\bibitem{idjadi2017integrated}
M.~H. Idjadi and F.~Aflatouni, ``Integrated pound- drever- hall laser
  stabilization system in silicon,'' {\em Nature communications}, vol.~8,
  no.~1, pp.~1--9, 2017.

\bibitem{joshi2016thermally}
C.~Joshi, J.~K. Jang, K.~Luke, X.~Ji, S.~A. Miller, A.~Klenner, Y.~Okawachi,
  M.~Lipson, and A.~L. Gaeta, ``Thermally controlled comb generation and
  soliton modelocking in microresonators,'' {\em Optics letters}, vol.~41,
  no.~11, pp.~2565--2568, 2016.

\end{thebibliography}

\medskip
\begin{footnotesize}

\noindent \textbf{Corresponding authors}: \href{mailto:igor.kudelin@colorado.edu}{igor.kudelin@colorado.edu} and \href{mailto:scott.diddams@colorado.edu}{scott.diddams@colorado.edu}

\noindent \textbf{Funding}: 
This research was supported by DARPA GRYPHON program (HR0011-22-2-0009) and NIST.

\noindent \textbf{Acknowledgments}: 
The authors thank Brian Long for the artistic illustration in Figure 5, and Kristina Chang and Nazanin Hoghooghi for comments on the manuscript. Commercial equipment and trade names are identified for scientific clarity only and does not represent an endorsement by NIST.

\end{footnotesize}

\end{document}